**Comments to "Asking Photons Where They Have Been"** [A.Danan, D. Farfurnik, S.Bar-Ad and L.Vaidman, Phys.Rev.Letters **111**, 240402 (2013)]

The letter by Danan *et al* [1] describes an interesting experiment using nested Mach-Zehnder interferometers (MZI) with vibrating mirrors to "tag" the photons. Here, I question some of the theoretical interpretations in [1].

One purpose of the experiment is to experimentally investigate a proposal put forward by Vaidman in [2], in which he suggests as "(a) criterion of the past of a quantum particle …. the weak trace it leaves". In agreement with Vaidman, I take the concept of a "weak trace" to mean a non-vanishing weak value of the projection operator onto (the relevant part of) the path of the particle. This seems a reasonable criterion; however, it must be applied with some care, in particular with due considerations to the fact that a weak value depends on both a pre-selected and a post-selected state.

Consider, e.g., a simple, well-balanced MZI, in which photons entering one of the ports of the input beam-splitter (BS) always end up in the bright port of the output BS, never in the dark port. However, from the absence of photons at the dark port, one may of course not conclude that there are no photons in the MZI arms.

A similar, but slightly more subtle effect occurs in the experiment in [1]. Its differential detection technique means that an above-noise signal in the detector occurs only if there is interference between the leading order contribution – the zeroth order in the small parameter $\delta/\Delta \approx 0.5 * 10^{-3}$ of [1] – to the detector arm amplitude and terms linear in $\delta/\Delta$. In that case, a detector signal proportional to $(\delta/\Delta)^2$ occurs, which in its turn means an effect proportional to $(\delta/\Delta)^4$ in the power-spectrum exhibited in the figures in [1]. If there is no such interference, only a linear term, the detector signal is heavily suppressed, in fact by a factor of (at least) $(\delta/\Delta)^2$, implying a suppression by a factor $(\delta/\Delta)^4$ in the power spectrum.

Thus, the differential detection technique does not register, above the noise level, linear terms unless they involve such interference. But in a similar way as for the simple MZI, one may not from the absence of a signal in the differential detector deduce the absence of photons in the arms of the nested MZI set-up: there could be – and indeed are, as a detailed calculation shows – photons to linear order in the amplitude that the detection technique does not detect. The authors' statements like "(t)he photons tell us that they have been in the parts of the interferometer through which they could not pass" cannot be upheld.

The experiment in [1] is analyzed using arguments based on weak values in the two-state vector formalism [3]; this formalism is completely equivalent to a formalism simply using (adequately time-evolved) pre- and post-selected states. The pre- and post-selected states used in [1] are given in its eq. (1). But the choice of the backward-evolving state $< \Phi |$ is not correct. In fact, the prescription in [3] for the backward-evolving state is to use the post-selected state and then evolve it backwards in time with (the inverse of) the actual time-evolution operator. This time-evolution process may be read off from figure 2(b) in [1], where there is no transition towards the nested MZI in the lowermost beam-splitter. The correct choice for $< \Phi |$ must therefore be the (backward-evolving) state in the C-arm, completely invalidating the conclusions of eq. (3) in [1]. Expressed in the two-state vector description of fig. 3 in [1], there should be no green, dashed line from the lower-most beam-splitter to the mirror F.

In sum, statements in [1] like "(t)he photons do not always follow continuous trajectories" and "… they never left the nested interferometer…" are not corroborated by a closer analysis of the experiment of Danan *et al*.

Bengt E Y Svensson
Department of Astronomy and
Theoretical Physics,
Lund, Sweden

**Acknowledgement**
I thank Ruth E Kastner for drawing my attention to the paper by Danan *et al*, and for many constructive comments.